\documentclass{elsart}
\usepackage{graphicx}
\usepackage{amssymb}

\begin{document}
\begin{frontmatter}
\title{Stylized Facts in Internal Rates of Return\\ 
on Stock Index and its Derivative Transactions}
\thanks[footnoteinfo]{Corresponding author.\\
Fax: +81-422-33-3286, E-mail: lukas@icu.ac.jp\\}

\author{Lukas Pichl,${}^{^{\mbox{\tiny 1,*}}}$ Taisei Kaizoji,${}^{^{\mbox{\tiny 2}}}$ and 
Takuya Yamano${}^{^{\mbox{\tiny 2}}}$ }
\address{%
${}^{^{\mbox{\rm \tiny 1}}}$ Division of Natural Sciences
}%
\address{%
${}^{^{\mbox{\rm \tiny 2}}}$ Division of Social Sciences\\
\hspace*{0mm}\\
International Christian University\\
Osawa 3-10-2, Mitaka, 181-8585 Japan
}%
\address{%
\rm Received 31 July 2006
}%

\begin{abstract}
Universal features in stock markets and their derivative markets are studied by 
means of probability distributions in internal rates of return on buy and sell 
transaction pairs. Unlike the stylized facts in log normalized returns, 
the probability distributions for such single asset encounters incorporate 
the time factor by means of the internal rate of return defined as the continuous compound interest. 
Resulting stylized facts are shown in the probability distributions derived 
from the daily series of TOPIX, S $\&$ P 500 and FTSE 100 index close values. 
The application of the above analysis to minute-tick data of NIKKEI 225 and its 
futures market, respectively,  reveals an interesting diffference in the behavior of 
the two probability distributions, in case  a threshold on the minimal duration
of the long position is imposed. It is therefore suggested that the probability 
distributions of the internal rates of return could be used for causality mining 
between the underlying and derivative stock markets. The highly specific discrete 
spectrum, which results from noise trader strategies as opposed to the smooth 
distributions observed for fundamentalist strategies in single encounter 
transactions may be also useful in deducing the type of investment strategy from trading 
revenues of small portfolio investors.       

\end{abstract}

\end{frontmatter}

\section{Introduction}

Stylized facts in market indicator data \cite{s1,s2,s3,s4,s5,s6,s7} have been reported, up to date, 
especially on the session-to-session basis, such as the fat tail law in the distribution 
of $R_t\equiv$log($P_i/P_{i-1}$) returns on various time scales and aggregation levels 
for data $P_i$ at ticks $i=1,..,N$. The reasons for this kind of statistics 
are well known: the values $P_i$ and $P_{i+1}$ should a priori correlate most strongly;
the relative factor $P_i/P_{i-1}$ is suitable for the analysis of market trends; the application
of the $\log$ function in addition brings symmetry to the two cases 
of indicator increase and decrease, while preserving the original meaning of small relative
market changes, $\log(1+x)\sim x$ for $|x|<<1$. Finally, it is well established 
that the histograms of $R$ values are practically symmetric for long-term data series 
with respect to the inversion of $R\rightarrow-R$, and may universally have 
fat tails relative to the normal distribution, whcih are known as the stylized facts. This constitutes a strong case
for studying indicator data in terms of probability distributions of normalized log
returns. Last but not least, the ratio $P_i/P_{i-1}-1$ is the profit
(or loss) from buying the asset at session $i$ and selling it at session $i+1$ 
without transaction costs. 

In this paper, we follow and extend the 
interpretation of $R_t$ as an elementary transaction (or function) for analyzing the aggregate
market data. While the standard indicator, $P_i/P_{i-1}-1$, fixes the single asset encounter
period $\Delta i=1$ and allows both for profit and loss, here we allow for a general $\Delta i<N$, determining this duration
of the asset encounter as the minimum $i´_0$ of all $i'>i$, for which a required {\sl internal rate
of return} is achieved. In the former approach, the computed log return values $R_t$ 
are partitioned into histogram bins to obtain the probability distribution; in the latter one, 
the probability distribution is obtained directly as the ratio of $i'_0$s, $i'_0<N$. This defines 
a transform over an interval of fixed internal rates of returns. Because of 
the exponential behavior of compound interesting, the size of the data sample has no 
effect on the distributions of internal rate of return (IRR) even for moderate 
values of $N$. Let us also note that in both cases, the $R$ and IRR values are obtained
ex post from index data, and therefore their economic interpretation implicitly
assumes no feedback between the virtual transactor and the market. The present approach
complements a variety of econometric, econophysics, statistical 
and artificial intelligence approaches in this field \cite{m1,m2,m3,m4}.    
               
Although the case for the use of the normalized log returns and IRR distributions 
has been presented above, it is worthwhile relating the independent variable of
both probability distributions, i.e. the return of a certain transaction, to the 
ultimate paradigm of market investment, the optimal transaction. The problem of optimal 
investment \cite{slanina} is also central to financial economics. Both the $R$ and IRR distributions
are time probabilities of return on a risky transaction with either two-component 
portfolio of money and the stock index, or multi-component portfolio of money 
and particular stock titles for index components. Depending on whether the transaction profit 
is reinvested or not, the geometric mean or the arithmetic mean of the expected
return of single encounter investment should be optimized \cite{bern}. The former
case of geometric mean is known as the Kelly criterion \cite{kelly}, which specifies that the optimal
single encounter ratio $x$ for one risky asset with return $b$ and probability $p$ 
is $x=p-(1-p)/b$, in order to maximize  the expectation value of the logarithm 
of the single encounter results, i.e. the economics' utility function. This is where 
the present single asset encounter formulation relates to Shannon's information 
entropy, and its generalized models.  
        
The above proposed IRR formulation ofthe  stylized facts in market indicator time series has, in fact, 
one more important motivation. While the log normalized returns build on the $[i-1,i]$
transaction pairs, the internal rate of return is defined for any $[i,i']$ transaction,
$0<i<i'<N$, not necessarily restricting to the preset fixed value in the IRR transform above. Thus, given 
a market and any sort of single-encounter formulated strategy, one obtains an IRR probability
spectrum, which can characterize both the market and the strategy.

The paper is organized as follows. Section 2 explains the theoretical rationale 
of the generalized stylized features as a transform in the internal rates of return.
The single encounter investment in the continuous interest model is applied to
the time series of TOPIX, S $\&$ P 500 and FTSE 100 data, for which the resulting IRR
transforms are derived and their stylized features are discussed. In Section 3, 
the characteristic spectrum of one particular single encounter strategy based 
on the comparison of short term and long term trends is shown. Section 4 demonstrates both 
formulations of the stylized facts on the minute-tick data from the NIKKEI 225
index, and its futures market, and discusses their causal relation in terms
of the IRR transform. We conclude with final remarks in Section 5.  

\section{Internal Rates of Return - Market Transform}

\begin{figure*}
\label{fig:f1}
\begin{center}
\includegraphics[width=0.9\hsize]{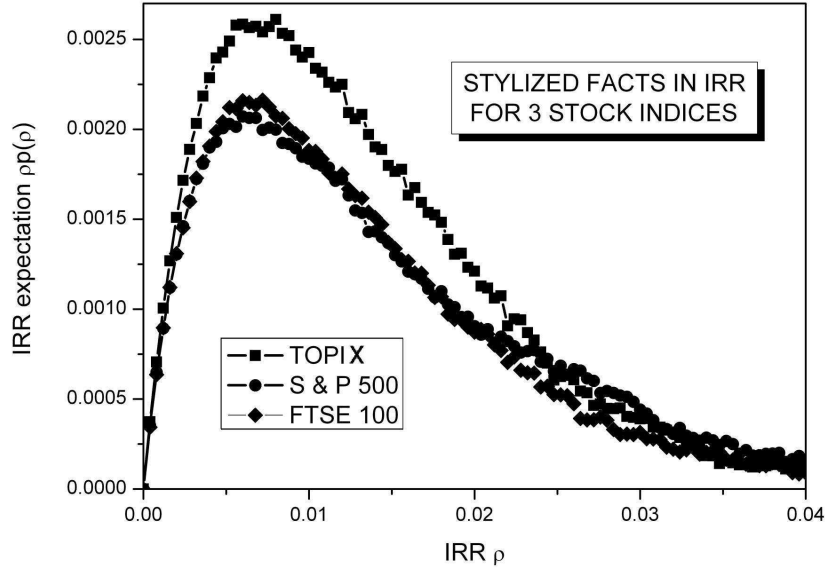}    
\caption{Stylized facts in the IRR transform for three stock indices in the period 
of 1995/09/01 to 2005/08/30 (daily quotes).}
\end{center}                                 
\end{figure*}

As mentioned above, the stylized facts in indicator time series [3-7] are usually studied by means of log-normalized returns, 
log($P_i/P_{i-1}$). Let us consider an investment on stock market index (which itself is one of the fundamental types of portfolios) 
for a variable period of $t$ sessions with a minimal required profit rate $r$ per session. In other words, 
\begin{equation}
\label{term}
0=C_t-\frac{C_0}{(1+r)^t}
\end{equation}
with an unknown period  $t$, cash flow $C_0$ equal to the buying price, and cash flow $C_t$ equal to the selling price.
The termination condition for the investment thus requires 
\begin{equation}
P_t\geq P_0(1+r)^t
\end{equation}
at the smallest possible time $t$ for which the above inequality holds. In economics terms, this would e.g. correspond
to the maximal implicit interest rate $r$ of reverse repo operation in the interval $<0,t>$. The required value of $r$ will 
be the basis of the IRR transform\footnote{Although the fixed $r$-value is a mathematical concept here, it also corresponds
to the fundamentalist strategy}. 
 
By using the compound discounting across market sessions, and the standard identity
$$
\lim_{n\rightarrow\infty}\left(1+\frac{a}{n}\right)^n=e^a,
$$ 
the termination condition in Eq. (\ref{term}) is equivalent to 
$$
P_t\geq P_0e^{\rho t},
$$
where $\rho $ is the IRR density in the continuous compound interest model, hereafter
the independent variable for the IRR transform. Let us note that in case of  $t=1$, 
the $\rho$ value  approximately reduces to the log-normalized return. 
For any fixed value of $\rho$, there is a certain probability $p(\rho)$ to 
realize this value in the market, which can be computed for instance 
by Monte Carlo simulation or a full scan of all data points, $i=1,..,N$, depending
on the size of data sample and the required accuracy. The boundary conditions
for $p(\rho)$ in standard data samples are $\rho\rightarrow 0 \Rightarrow p \rightarrow 1$,
and $\rho\rightarrow\infty \Rightarrow p \rightarrow 0$. For 
most actual market data, larger values of  $\rho$  imply a very rapid decay 
of $p(\rho)$,  because of the exponential explosion in the compound interest function, which cannot match the limited rates of return achievable in the real markets. 

The quantity $(\rho p(\rho)$ is the expected IRR, and must have a non-trivial local maximum at some point $\rho^*$. 
Since in general $t\neq 1$, the stylized facts in $I(\rho)\equiv \rho p(\rho)$ are different from 
the stylized facts in the log normalized returns, which would be $R(\rho)p(\rho)$.  
More than 30\% of the actual distributions dealt with here are found to  correspond to $\Delta t\geq 2$ events. 

Figure 1 shows such probability distributions for 2458 daily values of three stock indices, 
TOPIX, S $\&$ P 500 and FTSE 100 for a ten year period. It is interesting to note that
the optimal point $\rho^*$ coincides to high accuracy for all three distributions,
$\rho^*\sim 0.007$, an unexpected fact even in  case when the asymptotic 
power law exponents were similar. The close similarity of the S$\&$P 500 and FTSE 100
distributions, on the other hand, may be rather based on the close ties between
the U.S. and the U.K. stock markets. The properties of the IRR transform and its
generalizations will also be discussed in Section 4. 

The empirical distribution $I^*(\rho)$ for TOPIX in Fig. 1 serves as a reference case 
for the trend investing case discussed in the next section. 

\section{Internal Rates of Return - Strategy Transform}

\begin{figure*}
\label{fig:f2}
\begin{center}
\includegraphics[width=0.8\hsize]{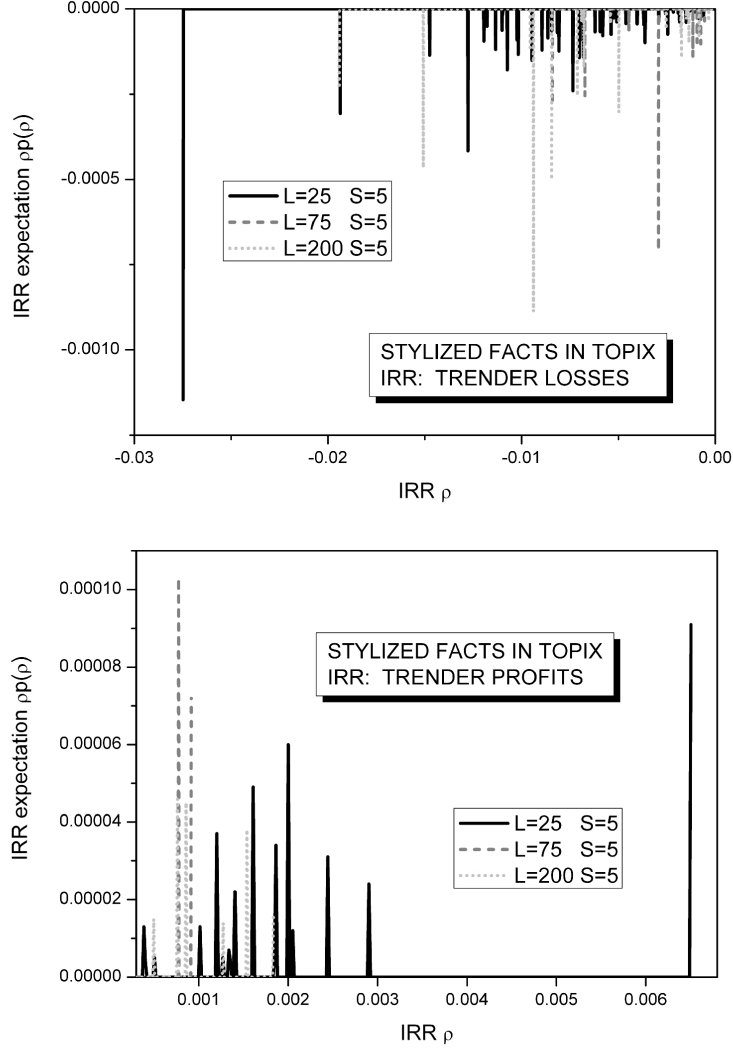}    
\caption{Discrete spectrum of the Golden Cross / Dead Cross strategy for
the TOPIX data between 1995/09/01 and 2005/08/30 (cf. Fig. 1 for the 
IRR transform). Upper panel: losses, lower panel: profits. The short-term
trend is taken as one week floating average (S=5); the curves
show the variance in the spectrum for middle to long term trends 
defined by 25, 75 and 200 data point averages.}
\end{center}                                 
\end{figure*}

Since the log-normalized returns do not allow for direct evaluation of 
nontrivial period single encounter strategies, it is interesting 
to study the spectral projection of such strategies, i.e. the I($\rho$) distribution, 
within the IRR transform, namely the difference from the reference
spectrum $I^{*}(\rho)$ (cf. the curve for TOPIX in Fig. 1).  This section discusses 
the possible benefits of such an approach, and uses the moving average
Dead Cross / Golden Cross trend strategy  for illustrations.

To address the noise trading, we adopt a buy signal / sell signal strategy, which is based on the comparison 
of short and middle term trends, which is often used as a reference by online brokers and elementary 
investment handbooks. This approach is based on the comparison of 
moving averages (MAs),
$$
P^{MA}(t)=\frac{1}{N}\sum_{n=0}^{n_{max}-1}P_{t-n},
$$
in the short-term ($n_{max}=S=5$) and middle-term ($n_{max}=L=25,75,200$) periods.

According to the model, a strong bullish market trend is signaled by an intersection of the 
rising short-term and middle-term curves, a signal to buy. A strong bearish market trend 
is signaled by an intersection of the falling short and middle term market curves, a 
signal to sell. These two turning points are routinely referred to as the Golden Cross 
and the Dead Cross, and are usually associated with analysts' recommendations 
to buy and sell, respectively. Although the Golden Cross point could be used in connection with 
some required IRR value, for instance, we assume the investment lasts from the buy to the
sell signal, which is consistent e.g. with the period from market bubble formation 
to its burst. Thus, provided the two signals originate on the opposite sides 
of a local maximum in the market, an investment between the two turning points 
may turn profitable. Since both cross points are based on relative criteria, there is 
no a priori guarantee that the resulting IRR density $I(\rho)$ from such transactions 
will in fact be positive, in contrast to the case of IRR market transform.

The strategy is formalized as follows. 
Let us denote the daily short-term and middle-term MA time series as
$\theta_i$ and $\lambda_i$, respectively, and their difference $\delta_i=\theta_i-\lambda_i$. 
The investment policy is then determined by the buy and sells signals 
arising in the market:
\begin{itemize} 
\item Buy: $\Delta_i\times\Delta_{i-1}<0$ (cross), $\lambda_i-\lambda_{i-1}>0$ (bullish long-term trend),
$\theta_i-\theta_{i-1}>0$ (bullish short-term trend) and $\theta_i+\lambda_{i-1}-(\theta_{i-1}+\lambda_i)>0$
(acceleration of the long-term bullish trend).
\item Sell: $\Delta_i\times\Delta_{i-1}<0$ (cross), $\lambda_i\times\lambda_{i-1}<0$ (bearish long-term trend),
$\theta_i-\theta_{i-1}<0$ (bearish short-term trend) and $\theta_i+\lambda_{i-1}-(\theta_{i-1}+\lambda_i)<0$
(acceleration of the long-term bearish trend).
\end{itemize}

By using the Monte Carlo computer simulation method, the initial day of investment $i_0$ 
is decided at random between the values $i=1$ and $N$, the size of the data set. 
Then the data are scanned for $i_0<i\leq N$ until the buy signal is found at some $i=i_b$.
The day $i_b$ corresponds to the start of the investment; the data are scanned again, 
$i_b<i\leq N$, for the reverse signal to terminate the investment at some $i=i_s$. 
The termination condition defines the respective $\rho$ resulting from each transaction
$k$ as 
$$
P_{i^{(k)}_s}/P_{i^{(k)}_b}=\exp(\rho^{(k)}(i^{(k)}_s-i^{(k)}_b)), 
$$
and its relative weight in the contribution to the IRR spectrum is $\rho^{(k)}$. 

Figure 2 shows the discrete spectrum for the 2458 daily values of the TOPIX index 
for three periods $L$ defining the middle-term trend. As the length of the middle
term increases, the frequency of cross events decreases, and the distribution 
becomes more singular. The spectrum $I(\rho)$ extends to the negative region of 
$\rho$. Although the data for the S $\&$ P 500 and FTSE 100 indices are not shown
in Fig. 2, their corresponding spectra are also discrete, and specific
for each market. However, because of the relatively infrequent cross point events 
in the data set, the statistics in Fig. 2 is rather qualitative. It is, however,
sufficient to support the discrete spectrum findings. 

\begin{figure*}
\label{fig:f3}
\begin{center}
\includegraphics[width=0.7\hsize]{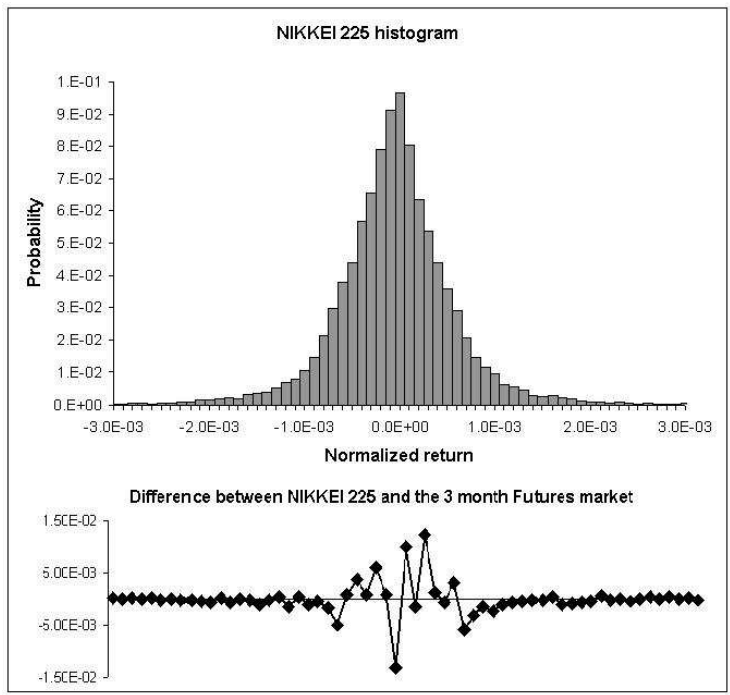}    
\caption{The histogram of log normalized returns of NIKKEI 225 (upper panel)
and the difference from the probability distribution in the futures market (lower panel).}
\end{center}                                 
\end{figure*}

\section{IRR Transform and Causality}

In this section, the advantages of the normalized log return and 
the IRR transform of  minute-tick data from NIKKEI 225 and its futures market 
are examined.
The data sample is taken between 9:01 on January 4th to 15:05 on June 8th, 2000,
which covers most of the first two quartal NIKKEI 225 futures
emissions in 2000. The log normalized return distribution for the NIKKEI 225
index, and the differences of the futures market from the underlying
index are shown in the upper and lower panel of Fig. 3, respectively. 
Within the statistic of this 29,810 point data sample, the difference
of the two $R-$distributions has more than 8 nodes within the FWHM region,
sized in the order of up to 10\% of NIKKEI 225 histogram. Although this is 
certainly an interesting result, there is no straightforward
way how to interpret the number of physically relevant nodes 
in the differential histogram.

\begin{figure*}
\label{fig:f4}
\begin{center}
\includegraphics[width=\hsize]{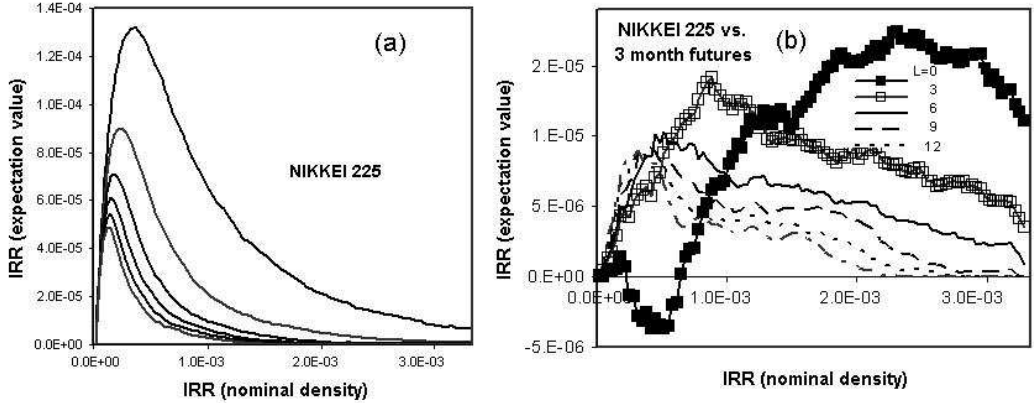}    
\caption{The IRR transform: (a) The $I(\rho)$ distributions of NIKKEI 225 index for minimal investment
periods of $T$=0, 3, 6, 9, 12, and 15 sessions. (b) The difference of $I(\rho)$ 
distributions between the NIKKEI 225 market and its derivative market.}
\end{center}                                 
\end{figure*}

Next, let us proceed to the IRR transform of the NIKKEI 225 index and its futures  market
data sample. In Fig. 4(a), the $I(\rho)$ distributions obtained with an imposed minimal
investment period $\tau$ are shown. Because of the compound interest, it becomes increasingly 
difficult to maintain a constant IRR density $\rho$ over a prolonged period; therefore 
the height of the distributions monotonously decreases with $\tau$ increasing in the 
series of 0, 3, 6, 9, 12 and 15 minutes. More interestingly, Fig. 4(b) shows the differences
of $I^{(\tau)}(\rho)$ between the index and its futures market for the two probability 
distribution $\tau$-series. Only in the case of $\tau=0$, there appear two nodes, with the maximum value 
in the negative direction approximately corresponding to the peak of the $I(\rho)$ distributions. 
For $\tau\geq 2$, all nodes disappear. In a different study, we have proposed 
an optimized symbolic analysis method for the alignment of the indicator time series \cite{symbol}.
Based on the alignment of the underlying index with the futures data series,
we suggest that the nodes for $\tau=0$ correspond to the lead-lag relation of the two markets within
this region, which is approximately $\Delta t\sim 1$. Together
with the finding in Fig. 4(b), it therefore appears that a series of the $I^{(\tau)}(\rho)$ 
for various values $\tau$ may be useful in lead-lag causality analysis of closely
related financial markets.

\section{Conclusion}

A framework for studying market statistics and its stylized facts based on the internal
rate of return from single transactions has been proposed and compared with the common
log normalized return approach. Based on the present formulation, an interesting stylized
fact was observed in the coinciding location of the optimal density of internal
rate of return, $\rho^*$, in the $I(\rho)$ probability distributions
of TOPIX, S $\&$ P 500 and FTSE 100 index data series on a ten year scale. The IRR
transform allows, in addition, for a spectral characteristics of single encounter
strategies, which was demonstrated on the TOPIX data series for several cases 
of trend-following strategies. The discrete $p(\rho)$ spectrum may serve as a fingerprint
of particular market strategy; it can also possibly allow for determining the type of 
investment strategy from the series of its returns on individual transactions. The
discrete spectrum feature may also present an interesting point of view on the possible 
role of moving averages in the stylized facts in  numerical time series. The 
IRR transform has been further generalized by introducing a minimal interval $\tau$
on compound interest time period, which screens the $I(\rho)$ spectrum in a monotonous
way. This generalization allowed for indirect identification of the lead-lag
relation in the minute-tick series of NIKKEI 225 and its derivative market in
the first two quartals of the year 2000.    

\section*{Acknowledgments}
The authors would like to acknowledge a partial support by the Japan
Society for the Promotion of Science (JSPS) and Academic Frontier
Program of the Japanese Ministry of Education, Culture, Sports, Science
and Technology (MEXT).

\end{document}